\documentclass[conference]{IEEEtran}
\def\BibTeX{{\rm B\kern-.05em{\sc i\kern-.025em b}\kern-.08em
    T\kern-.1667em\lower.7ex\hbox{E}\kern-.125emX}}
\usepackage{amsmath}
\usepackage{amsthm}
\usepackage{amsfonts}
\usepackage{amssymb}
\usepackage{graphicx}
\usepackage{cite}
\usepackage{subfigure}
\usepackage{balance}
\usepackage{algorithm}
\usepackage{algorithmic}
\usepackage{enumerate}
\usepackage{color}
\usepackage{array}
\usepackage{indentfirst}
\usepackage{multirow}
\usepackage{booktabs}
\usepackage{color}
\usepackage{slashbox}
\usepackage{diagbox}
\usepackage{threeparttable}
\usepackage[svgnames,table]{xcolor}
\usepackage{bm}
\usepackage{bbm}
\usepackage{subfigure}
\usepackage{epstopdf}

\makeatletter

\newcommand{\Rmnum}[1]{\expandafter\@slowromancap\romannumeral #1@}
\makeatother

\begin{document}
 	\title{{Deep Learning-empowered Predictive Precoder Design for OTFS Transmission in URLLC}}

\author{Chang Liu$^{\ast}$, Shuangyang Li$^{\dag}$, Weijie Yuan$^{\ddag}$, Xuemeng Liu$^{\P}$, and Derrick Wing Kwan Ng$^{\S}$ \\
$^{\ast}$Department of Electronic and Information Engineering, The Hong Kong Polytechnic University, Hong Kong, China \\
$^{\dag}$Electrical Engineering and Computer Science Department, Technische Universit{\"a}t Berlin, Germany \\
$^{\ddag}$Department of Electronic and Electrical Engineering, Southern University of Science and Technology, China \\
$^{\P}$School of Electrical and Information Engineering, University of Sydney, Sydney, Australia \\
$^{\S}$School of Electrical Engineering and Telecommunications, University of New South Wales, Sydney, Australia
%Email: $^{\S}$\{chang.liu19, w.k.ng\}@unsw.edu.au, shuangyang.li@uwa.edu.au, \\
%$^{\ddag}$yuanwj@sustech.edu.cn, $^{\S}$xuemeng.liu@sydney.edu.au

}

\maketitle

\begin{abstract}
To guarantee excellent reliability performance in ultra-reliable low-latency communications (URLLC), pragmatic precoder design is an effective approach.
However, an efficient precoder design highly depends on the accurate instantaneous channel state information at the transmitter (ICSIT), which however, is not always available in practice.
To overcome this problem, in this paper, we focus on the orthogonal time frequency space (OTFS)-based URLLC system and adopt a deep learning (DL) approach to directly predict the precoder for the next time frame to minimize the frame error rate (FER) via implicitly exploiting the features from estimated historical channels in the delay-Doppler domain.
By doing this, we can guarantee the system reliability even without the knowledge of ICSIT.
To this end, a general precoder design problem is formulated where a closed-form theoretical FER expression is specifically derived to characterize the system reliability.
Then, a delay-Doppler domain channels-aware convolutional long short-term memory (CLSTM) network (DDCL-Net) is proposed for predictive precoder design.
In particular, both the convolutional neural network and LSTM modules are adopted in the proposed neural network to exploit the spatial-temporal features of wireless channels for improving the learning performance.
Finally, simulation results demonstrated that the FER performance of the proposed method approaches that of the perfect ICSI-aided scheme.
\end{abstract}

\section{Introduction}
Ultra-reliable and low-latency communications (URLLC) is one of the key enabling technologies in realizing the upcoming beyond fifth-generation (B5G) and sixth-generation (6G) \cite{park2022extreme}.
In URLLC, a block error rate of lower than $0.001 \%$ and a latency of shorter than 1 milliseconds (ms) are usually required to facilitate ultra-high reliability and low latency, which promotes numerous emerging applications \cite{siqin2022platform, hu2021robust, cai2022resource, li2022many, liu2020deepresidualconference}.
Thus, the research on URLLC has attracted vast attention from both academia and industry.

Recently, various effective and efficient schemes have been proposed for realizing URLLC.
For instance, the authors in \cite{She2018cross_layer} proposed an URLLC-oriented cross-layer optimization method taking into consideration of both the transmission and queueing delays.
In addition, \cite{Chang2019optimizing} studied the URLLC and control subsystems by developing a joint resource optimization scheme to maximize the overall system performance.
It is worth noting that most URLLC schemes focused on the signal processing in the time-frequency (TF) domain, which incurs some inherent communication unfriendly properties \cite{tse2005fundamentals}, such as limited path lifetime, severe delay and Doppler spread, etc.
To overcome these drawbacks, orthogonal time frequency space (OTFS) modulation \cite{Hadani2017orthogonal, Li2020performance} which exploits the quasi-static, sparse, and compact delay-Doppler (DD) domain for advanced signal processing, has been proposed as a promising solution to the implementation of URLLC \cite{Zhiqiang_magzine}.
For example, \cite{Raviteja2019effective} studied the diversity of uncoded OTFS transmission and the results found that OTFS can almost achieve the full channel diversity.
However, the achieved excellent OTFS error performance relies on the maximum-likelihood detection with a prohibitively high computational complexity \cite{li2022novel}.
As an alternative, low-complexity detection methods, e.g., the zero-forcing (ZF) and minimum mean square error (MMSE) detectors, can be adopted to improve the system practicability.
Nevertheless, these practical methods only have a very limited performance in signal detection.
To further improve the system bit error rate performance, a precoder scheme was studied for OTFS transmission in \cite{HuangQin2022ComLet}. However, this precoder scheme requires the knowledge of perfect instantaneous channel state information at the transmitter (ICSIT), which is practically impossible for the low latency communications.
Indeed, the acquisition of ICSIT hinders the design of a practical and efficient precoder for URLLC systems.

\begin{figure*}
\centering
\includegraphics[width=0.6\linewidth]{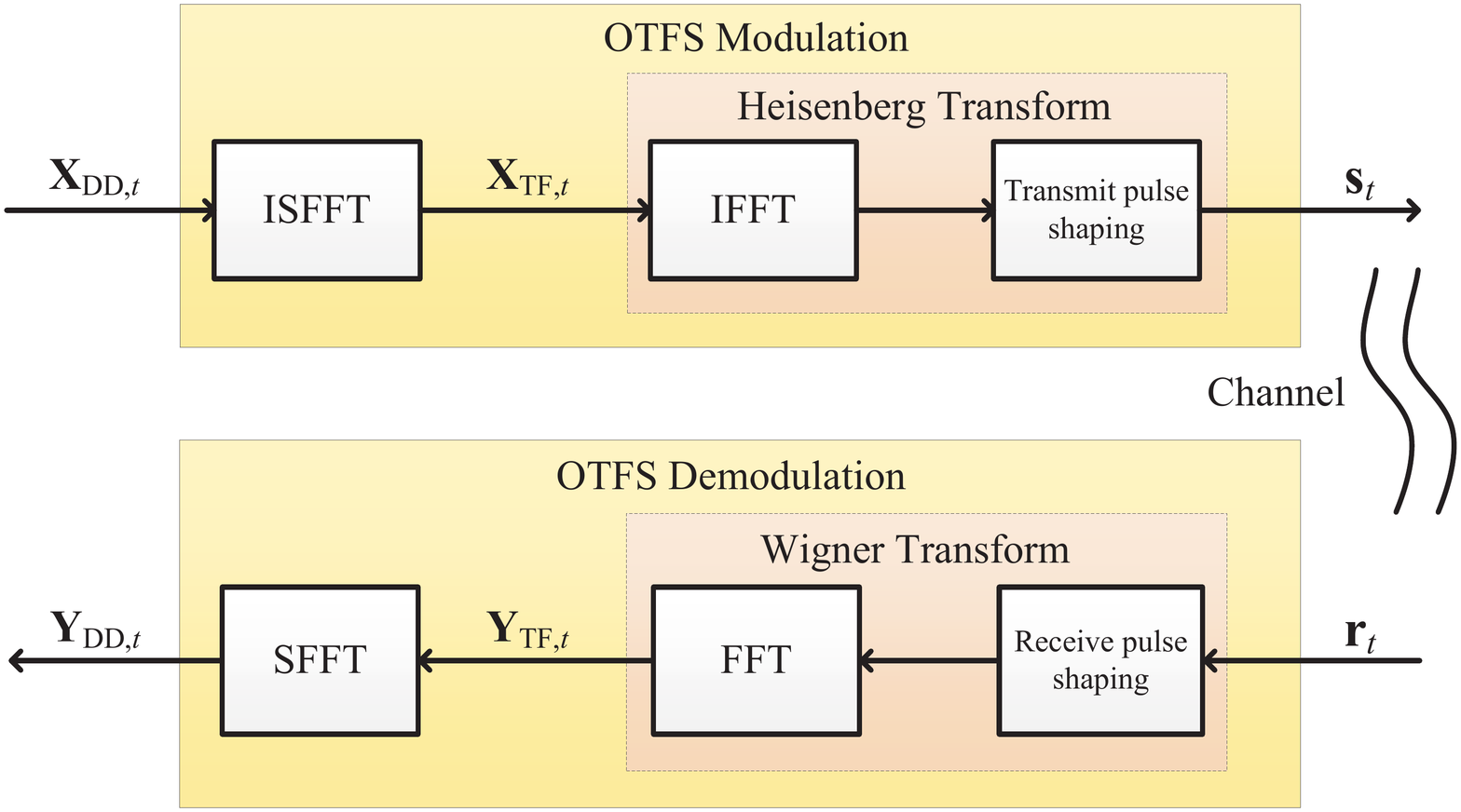}
\caption{The considered OTFS-enabled URLLC system.}\label{scenario}
\end{figure*}

Note that if the precoder in the next period can be well predicted in advance, the system performance in the subsequent period can be guaranteed even without the need for ICSIT acquisition.
On the other hand, recently, the recurrent neural network (RNN) \cite{goodfellow2016deep, o2017introduction, xie2019activity, yuan2020learning}, which adopts a recurrent feedback mechanism to extract temporal dependencies for predicting the dynamic behaviors, has proven its efficiency in predictive tasks.
Inspired by this, in this paper, we adopt a deep learning (DL) approach \cite{wang2017deep, liu2020deeptransfer, dai2020deep, liu2022learning, liu2022deepresidual, xie2020unsupervised, xie2020deep}, i.e., an RNN-based scheme to exploit features from the historical DD domain channels to design an predictive precoder for the next period, which can further improve the system reliability without the knowledge of ICSIT.
To this end, we first derive a closed-form theoretical expression of the frame error rate (FER) to characterize the system reliability and thus formulate a general predictive precoder problem to minimize the FER subject to the power constraint at the transmitter.
Then, to solve the formulated problem, we propose a DD domain channels-aware convolutional long short-term memory (CLSTM) network (DDCL-Net), where both the convolutional neural network (CNN) and LSTM modules are successively adopted to exploit the spatial-temporal features of historical DD domain channels to improve the learning performance for predictive precoder design.
Finally, simulation results demonstrated that the FER performance of the proposed method can approach the lower bound achieved by a DL scheme with perfect ICSI at both the transmitter and the receiver.

\emph{Notations:}
Superscripts $H$ and $T$ are adopted to represent the conjugate transpose and transpose, respectively.
$\mathbb{R}$ and $\mathbb{C}$ represent the sets of real and complex numbers, respectively.
$\mathbb{N}_1$ is used to denote the natural number set.
$ \otimes $ represents the kronecker product.
${{{\bf{F}}_N}}$ and ${{{\bf{I}}_M}}$ represent the discrete Fourier transform (DFT) matrix with the size of $N\times N$ and the identity matrix with the size of $M\times M$, respectively.
$[a]_b$ represents the modulo operation of $a$ modulo $b$.
$\textrm{vec}(\cdot)$ is used to represent the vectorization operation and $\textrm{mat}_{MN}(\cdot)$ denotes the generation of a matrix with a size of $M \times N$ based on a vector with a size of $MN \times 1$.
$\textrm{diag}(\mathbf{a})$ represents the generation of a diagonal matrix based on the diagonal elements of $\mathbf{a}$.
${\mathcal{CN}}( \bm{\mu},\mathbf{\Sigma} )$ is used to represent a circularly symmetric complex Gaussian (CSCG) distribution with a mean vector of $\bm{\mu}$ and a covariance matrix of $\mathbf{\Sigma}$.
$U(a,b)$ represents a uniform distribution with the range of $[a,b]$.
$\|\cdot\|$ and $\|\cdot\|_F$ are adopted to denote the $\ell_2$-norm and the Frobenius norm of a vector and a matrix, respectively.
$|\cdot|$ represents the absolute value of a complex-valued number.
$\Pi_{k=1}^K x_k$ represents the product of $K$ numbers $x_1,x_2,\cdots, x_K$.
$\mathbb{E}\{\cdot\}$ is the statistical expectation operation.
$\mathrm{Re}\{\cdot\}$ and $\mathrm{Im}\{\cdot\}$ denote the real and imaginary parts of a complex-valued input, respectively.
In addition, $\mathbf{X}(a:b,:)$ denotes the part of a matrix $\mathbf{X}$ from the $a$-th row to the $b$-th row.

\begin{figure*}
\centering
\includegraphics[width=0.6\linewidth]{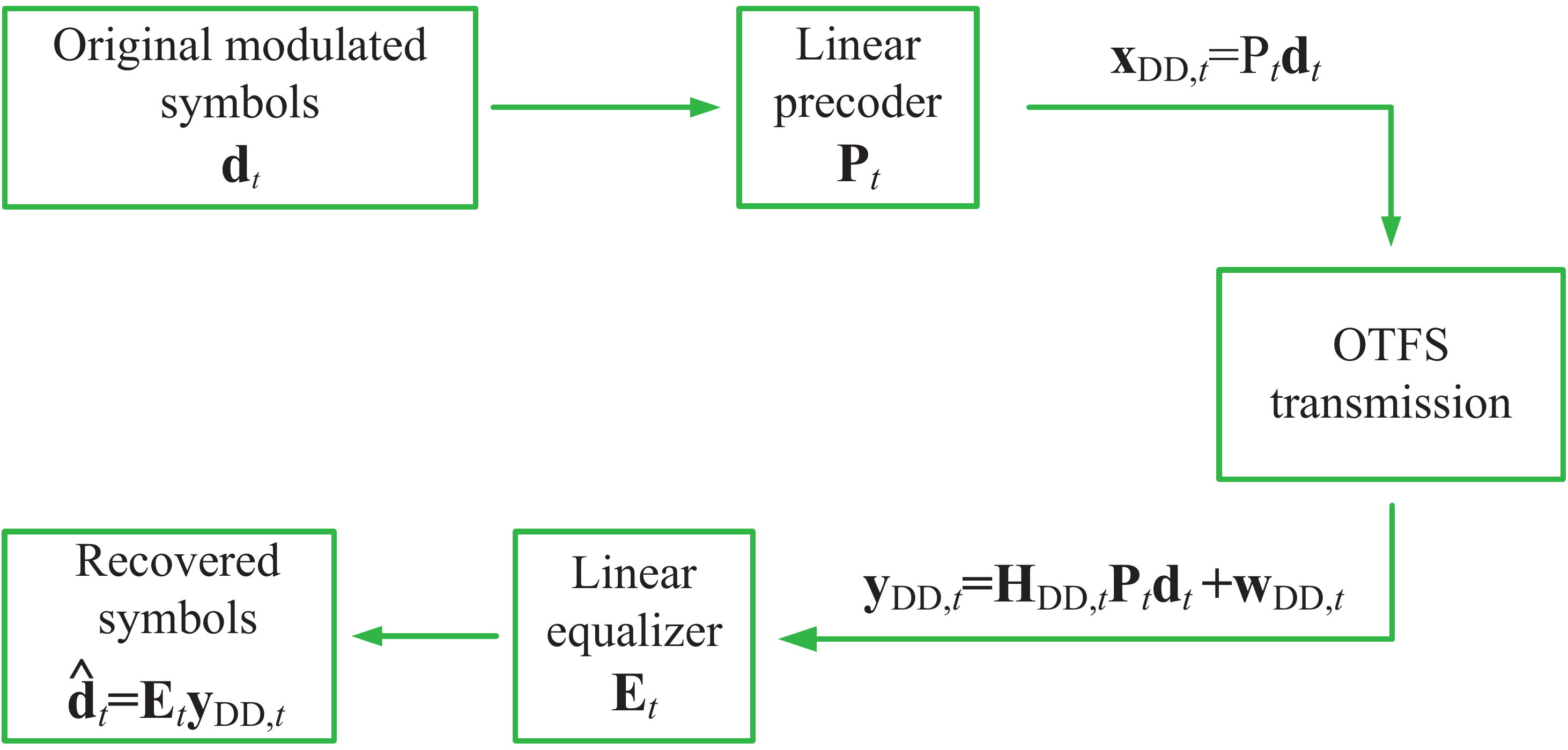}
\caption{The precoder-based OTFS transmission for URLLC.}
\label{precoder_model}
\centering
\end{figure*}

\section{System Model}

As shown in Fig. \ref{scenario}, in this paper, we consider an OTFS-enabled URLLC system \cite{li2021cross}, which consists of one single-antenna transmitter and one single-antenna receiver.
In addition, we assume that there are $M$ subcarriers and $N$ time slots, where the subcarrier spacing and the time slot duration are $1/T$ and $T$, respectively.
In particular, we consider a time-varying scenario and focus on one frame for study, where the channel parameters are constant within each frame and changes with different frames \cite{tse2005fundamentals}.
For ease of presentation, we adopt subscript $t$ to represent an arbitrary frame $t$ in the following.
Denote by $\mathbf{x}_{\mathrm{DD},t} \in \mathbb{C}^{MN \times 1} $ the DD domain symbol vector at frame $t$, where each element is uniformly generated from an energy-normalized constellation set $\mathbb A$.
The associated DD domain symbol matrix can be expressed as ${\bf X}_{\mathrm{DD},t} = \textrm{mat}_{MN}(\mathbf{x}_{\mathrm{DD},t}) \in \mathbb{C}^{M \times N}$.
According to \cite{Hadani2017orthogonal}, by exploiting the inverse symplectic finite Fourier transform (ISFFT), we can exploit the Heisenberg transform to obtain the TF domain symbol matrix $\mathbf{X}_{\mathrm{TF},t}$, which is given by
\begin{equation}\label{}
  \mathbf{X}_{\mathrm{TF},t} = \mathbf{F}_{M}{\bf X}_{\mathrm{DD},t}\mathbf{F}_{N}^H.
\end{equation}
Thus, we can express the time domain OTFS symbol vector at frame $t$, denoted by $\mathbf{s}_t \in \mathbb{C}^{MN \times 1}$, as
\begin{align}
{\mathbf{s}_t} = \left( {{\bf{F}}_N^{\rm{H}} \otimes {{\bf{I}}_M}} \right){{\bf{x}}_{{\rm{DD}}}}.
\end{align}
For ease of study, a DD deterministic channel is adopted for the considered system, where both the Doppler and delay responses of the underlying physical scatterers are assumed to be static over each signal transmission.
In this case, the time domain channel matrix is given by \cite{Raviteja2018interference}
\begin{align}
{\bf{H}}_{{\rm{T}},{t}} \buildrel \Delta \over = \sum\limits_{p = 1}^P {{h_{p,t}}{e^{ - j2\pi \frac{{{k_{p,t}}{l_{p,t}}}}{{MN}}}}{{\bm \Delta} ^{{k_{p,t}}}}{{\bf{\Pi }}^{{l_{p,t}}}}}. \label{H_T_t}
\end{align}
Here, $h_{p,t}$, ${l_{p,t}}$, and ${k_{p,t}}$ denote the complex fading coefficient, the delay index, and the Doppler index at frame $t$ with respect to (w.r.t.) the $p$-th resolvable path, respectively.
Also,
\begin{equation}
{{\bf{ \Pi }}}= {\left[ {\begin{array}{*{20}{c}}
0& \cdots &0&1\\
1& \ddots &0&0\\
 \vdots & \ddots & \ddots & \vdots \\
0& \cdots &1&0
\end{array}}
\right]} \in \mathbb{R}^{MN \times MN}
\end{equation}
and ${\bm \Delta}={\rm{diag}}( {1,{e^{j2\pi \frac{1}{{MN}}}},...,{e^{j2\pi \frac{{MN - 1}}{{MN}}}}} ) \in \mathbb{C}^{MN \times MN}$ denote the permutation matrix and the phase rotating matrix, respectively.
In particular, we adopt a channel offset model to characterize the time-varying property of the wireless environment, where ${{\bf{H}}_{\mathrm{T},t}}$ varies with frames and remains constant within each frame.
In this case, the delay and Doppler indices can be characterized by
\begin{equation}\label{}
  l_{p,t} = l_{p,t-1} + \epsilon_{p,t}
\end{equation}
and
\begin{equation}\label{}
  k_{p,t} = k_{p,t-1} + \varepsilon_{p,t},
\end{equation}
respectively. Here, $\epsilon_{p,t} \sim U(\epsilon_{\min},\epsilon_{\max})$ represents the delay index-offset at frame $t$ w.r.t. the $p$-th resolvable path, where $\epsilon_{\min}$ and $\epsilon_{\max}$ are used to represent the minimum and maximum offsets of the delay index, respectively.
In addition, the Doppler index-offset at frame $t$ w.r.t. the $p$-th resolvable path is denoted by $\varepsilon_{p,t} \sim U(\varepsilon_{\min},\varepsilon_{\max})$, where $\varepsilon_{\min}$ and $\varepsilon_{\max}$ are the minimum and maximum Doppler index-shifts, respectively.
Furthermore, to characterize the time-varying channel fading, we adopt a first-order complex Gauss-Markov model, which is given by \cite{nasir2019multi}
\begin{equation}\label{}
  h_{p,t} = \rho h_{p,t-1} + \sqrt{1 - \rho^2} \vartheta_{p,t}.
\end{equation}
Here, $\rho \in [0,1]$ is the parameter to characterize the correlation.
$\vartheta_{p,t} \sim \mathcal{CN}(0,1/P)$ denotes a CSCG random variable to characterize the time-varying change.

Based on the above discussion, the received time domain signal after removal of cyclic prefix (CP) can be formulated as
\begin{align}
{\mathbf{r}_t} = {{\mathbf{H}}_{\mathrm{T},t}}{\mathbf{s}_t} + {\mathbf{w}_t}. \label{TD_model_vec}
\end{align}
where ${\mathbf{r}_t} \in \mathbb{C}^{MN \times 1} $ and $\mathbf{w}_t \in \mathbb{C}^{MN \times 1}$ denote the received signal vector and the CSCG noise vector, respectively.
According to Fig. \ref{scenario}, by exploiting the Wigner transform and the SFFT, we can obtain the received DD domain symbol vector, i.e.,
\begin{equation}
{{\bf{y}}_{{\rm{DD}},{t}}} = \left( {{{\bf{F}}_N} \otimes {{\bf{I}}_M}} \right){\bf{r}}_t = {{\bf{H}}_{{\rm{DD}},{t}}}{{\bf{x}}_{{\rm{DD},{t}}}} + {{\bf{w}}_{{\rm{DD}},{t}}}. \label{DD_model_vec}
\end{equation}
Here, ${{\bf{y}}_{{\rm{DD}},{t}}} = {\rm{vec}}\left( {{{\bf{Y}}_{{\rm{DD}},{t}}}} \right) \in \mathbb{C}^{MN \times 1}$, where ${{{\bf{Y}}_{{\rm{DD}},{t}}}} \in \mathbb{C}^{M \times N}$ denotes the received DD domain symbol matrix.
${{\bf{w}}_{{\rm{DD}},{t}}} = \left( {{{\bf{F}}_N} \otimes {{\bf{I}}_M}} \right) \mathbf{w}_t \sim \mathcal{CN}(\mathbf{0},\sigma^2\mathbf{I}_{MN})$ represents the DD domain CSCG noise vector with $\sigma^2$ being the corresponding noise variance.
In addition, ${{\bf{H}}_{{\rm{DD}},{t}}} \in \mathbb{C}^{MN \times MN} $ represents the DD domain channel matrix at frame $t$, which can be formulated as
\begin{equation}
{{\bf{H}}_{{\rm{DD}},{t}}} = \left( {{{\bf{F}}_N} \otimes {{\bf{I}}_M}} \right){{\bf{H}}_{{\rm{T}},{t}}}\left( {{\bf{F}}_N^{\rm{H}} \otimes {{\bf{I}}_M}} \right).
\label{H_DD}
\end{equation}

\begin{figure*}[t]
  \centering
  \includegraphics[width=0.6\linewidth]{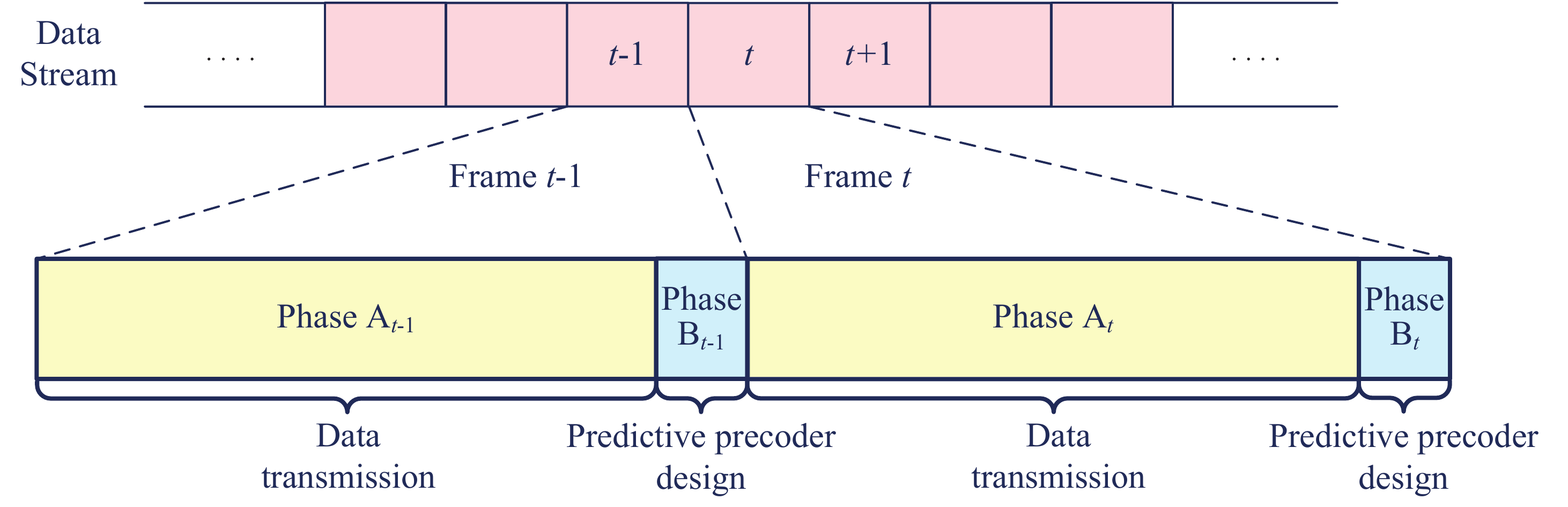}
  \caption{The proposed predictive precoder-based transmission protocol for the considered OTFS-enabled URLLC system.}\label{Fig:protocol_structure}\vspace{-0.2cm}
\end{figure*}

\section{Problem Formulation}
In this section, we first develop a predictive precoder-based transmission protocol and then derive the problem formulation for the predictive precoder design in the considered OTFS-enabled URLLC system.
The details will be introduced as follows.

\subsection{Predictive Precoder-based Transmission Protocol}
To further improve the system reliability, a precoder scheme can be added to the OTFS transmission, as shown in Fig. \ref{precoder_model}.
Given a modulated symbols $\mathbf{d}_t \in \mathbb{C}^{K \times 1}$ with $\mathbb{E}[\mathbf{d}_t\mathbf{d}_t^H] = \mathbf{I}_{K}$, we can exploit the precoder matrix at frame $t$, denoted by $\mathbf{P}_t \in \mathbb{C}^{MN \times K}$ to obtain the precoded symbols $\mathbf{x}_{{\mathrm{DD}},t}$, i.e.,
\begin{equation}\label{x_DD_t}
  \mathbf{x}_{{\mathrm{DD}},t} = \mathbf{P}_t\mathbf{d}_t.
\end{equation}
Thus, the received symbol vector can be expressed as
\begin{equation}\label{}
  \mathbf{y}_{{\mathrm{DD}},t}=\mathbf{H}_{{\mathrm{DD}},{t}}\mathbf{P}_t\mathbf{d}_t + {{\bf{w}}_{{\rm{DD}},{t}}}.
\end{equation}
Then, by adopting a classical MMSE equalizer, which is given by \cite{tse2005fundamentals}
\begin{equation}\label{E_t}
  \mathbf{E}_t = \left( \sigma^2 \mathbf{I}_{MN} + \mathbf{P}_t^H\hat{\mathbf{H}}_{{\mathrm{DD}},{t}}^H\hat{\mathbf{H}}_{{\mathrm{DD}},{t}}\mathbf{P}_t \right)^{-1}\mathbf{P}_t^H\hat{\mathbf{H}}_{{\mathrm{DD}},{t}}^H,
\end{equation}
we can finally recover the transmitted symbol at the receiver, i.e.,
\begin{equation}\label{d_hat}
  \hat{\mathbf{d}}_t = \mathbf{E}_t \mathbf{y}_{{\mathrm{DD}},t}.
\end{equation}
In this case, we can calculate the received signal-to-interference-plus-noise ratio (SINR) w.r.t. the $k$-th symbol at frame $t$, which can be formulated as
\begin{equation}\label{SINR}
\begin{aligned}
  &\mathrm{SINR}_{k,t} \\
  & = \frac{|[\mathbf{E}_t\mathbf{H}_{{\mathrm{DD}},{t}}\mathbf{P}_t]_{k,k}|^2}
  {|[\mathbf{E}_t\mathbf{H}_{{\mathrm{DD}},{t}}\mathbf{P}_t]_{k,:}-[\mathbf{E}_t\mathbf{H}_{{\mathrm{DD}},{t}}\mathbf{P}_t]_{k,k}|^2 + \sigma^2|[\mathbf{E}_t]_{k,:}|^2}.
\end{aligned}
\end{equation}
Given the SINR at the receiver, according to \cite{palomar2005minimum}, we can then derive the symbol error rate (SER) w.r.t. the $k$-th symbol in frame $t$ (under a Gray encoding scheme), i.e.,
\begin{equation}\label{SER}
  \mathrm{SER}_{k,t} \approx \alpha\,\mathrm{erfc} \left(\sqrt{\beta\mathrm{SINR}_{k,t}}\right),
\end{equation}
where $\alpha = \left(2 - 2/\sqrt{M_{\mathrm{Mod}}}\right)/\log_2 M_{\mathrm{Mod}}$, $\beta = 3/\left(2\sqrt{M_{\mathrm{Mod}}} - 2 \right)$, and $\mathrm{erfc}(\cdot)$ denotes the complementary error function.
Based on (\ref{SER}), we can finally derive the theoretical expression of FER at frame $t$ for the considered URLLC system, which is given by \begin{equation}\label{FER}
  \mathrm{FER}_t = 1 - \Pi_{k=1}^K(1 - \mathrm{SER}_{k,t}).
\end{equation}

\newcounter{mytempeqncnt1}
\begin{figure*}[!t]
\normalsize
\setcounter{mytempeqncnt1}{\value{equation}}
\setcounter{equation}{17}
{\begin{equation}\label{f_FER}
f_{\mathrm{FER}}(\mathbf{H}_{{\mathrm{DD}},{t}},\mathbf{P}_t) \triangleq 1 - \Pi_{k=1}^K\left(1 - \alpha \,  \mathrm{erfc}\left(\sqrt{\frac{\beta|[\mathbf{E}_t\mathbf{H}_{{\mathrm{DD}},{t}}\mathbf{P}_t]_{k,k}|^2}
  {|[\mathbf{E}_t\mathbf{H}_{{\mathrm{DD}},{t}}\mathbf{P}_t]_{k,:}-[\mathbf{E}_t\mathbf{H}_{{\mathrm{DD}},{t}}\mathbf{P}_t]_{k,k}|^2 + \sigma^2|[\mathbf{E}_t]_{k,:}|^2}}\right)\right).
\end{equation}}
\setcounter{equation}{\value{mytempeqncnt1}}
\hrulefill
\end{figure*}
\setcounter{equation}{18}

According to (\ref{x_DD_t})-(\ref{FER}), we can find that the design of $\mathbf{P}_t$ depends on accurate $\mathbf{H}_{{\mathrm{DD}},{t}}$ at the transmitter, which however, is not always available in URLLC since the system cannot grant enough time for uplink channel estimation at the transmitter.
As an alternative, if the required precoder can be designed in advance, we can guarantee the system FER performance in the subsequent period without the acquisition of the ICSIT.
Inspired by this, we develop a predictive transmission protocol in Fig. \ref{Fig:protocol_structure}.
For each frame $t$, there are two cascaded phases, i.e., Phase A$_t$: Data transmission and Phase B$_t$: Predictive precoder design.
In Phase A$_t$, based on the predictive precoder obtained from Phase B$_{t-1}$, the transmitter can directly execute the data transmission process.
Subsequently, in Phase B$_{t}$, one can exploit the estimated historical channels\footnotemark\footnotetext{These historical channels can be estimated by leveraging the receiver-to-transmitter feedback link.}, denoted by  $\hat{\mathbf{H}}_{{\mathrm{DD}},t-1}, \hat{\mathbf{H}}_{{\mathrm{DD}},t-2}, \cdots, \hat{\mathbf{H}}_{{\mathrm{DD}},t-\tau}$, to exploit the temporal dependency of the time-varying channels to design the precoder for frame $t+1$.

\subsection{Predictive Precoder Problem Formulation}
According to the proposed protocol and (\ref{SINR})-(\ref{FER}), we can formulate the problem for predictive precoder, i.e., $\forall t \in \mathcal{T} = \{t| t \geq \tau + 1, t \in \mathbb{N}_1 \}$, minimizing the statistical expectation of FER theoretical expression of frame $t$ subject to the power constraint.
Based on (\ref{SINR}) and (\ref{SER}), we can rewrite (\ref{FER}) as a function of $\mathbf{H}_{\mathrm{DD},{t}}$ and ${\mathbf{P}}_t$, as derived in (\ref{f_FER}) at the top of next page.
Thus, the predictive precoder problem for the considered OTFS-enabled URLLC can be formulated as
\begin{align}
(\mathcal{P}):~\mathop{\max}\limits_{{\mathbf{P}}_t } ~ &\mathbb{E}_{\mathbf{H}_{\mathrm{DD},{t}}|\mathcal{H}_{\mathrm{DD},{t}}^{\tau}}
\left\{f_{\mathrm{FER}}(\mathbf{H}_{{\mathrm{DD}},{t}},{\mathbf{P}}_t)
\right\}  \label{PF_objective} \\
\mathrm{s.t.}~&\|\mathbf{P}_t\|_F^2\leq P_0, \forall t \in \mathcal{T} . \label{PF_power}
\end{align}
Here, ${\mathbf{P}}_t \in \mathbb{C}^{MN \times K}$ denotes the predictive precoder to be optimized, as defined in (\ref{x_DD_t}).
$\mathbb{E}_{\mathbf{H}_{\mathrm{DD},{t}}|\mathcal{H}_{\mathrm{DD},{t}}^{\tau}}\{\cdot\}$ represents the statistical expectation of $\mathbf{H}_{\mathrm{DD},{t}}$ given the estimated historical channels $\mathcal{H}_{\mathrm{DD},{t}}^{\tau} \triangleq [\hat{\mathbf{H}}_{\mathrm{DD},{t-1}},\hat{\mathbf{H}}_{\mathrm{DD},{t-2}},\cdots,\hat{\mathbf{H}}_{\mathrm{DD},{t-\tau}}]$.
(\ref{PF_power}) is the power constraint with the maximum power budget $P_0$.
Note that it is very challenging to solve the formulated problem $(\mathcal{P})$ since it is intractable to directly derive the closed-form expression of the objective function.
On the other hand, the dedicated objective function is non-convex w.r.t. ${\mathbf{P}}_t$, thus it is quite challenging to find effective solutions.
To overcome these issues, in the next section, we will adopt a data-driven DL approach to address the formulated problem for predictive precoder design.

\begin{figure*}[t]
  \centering
  \includegraphics[width=0.97\linewidth]{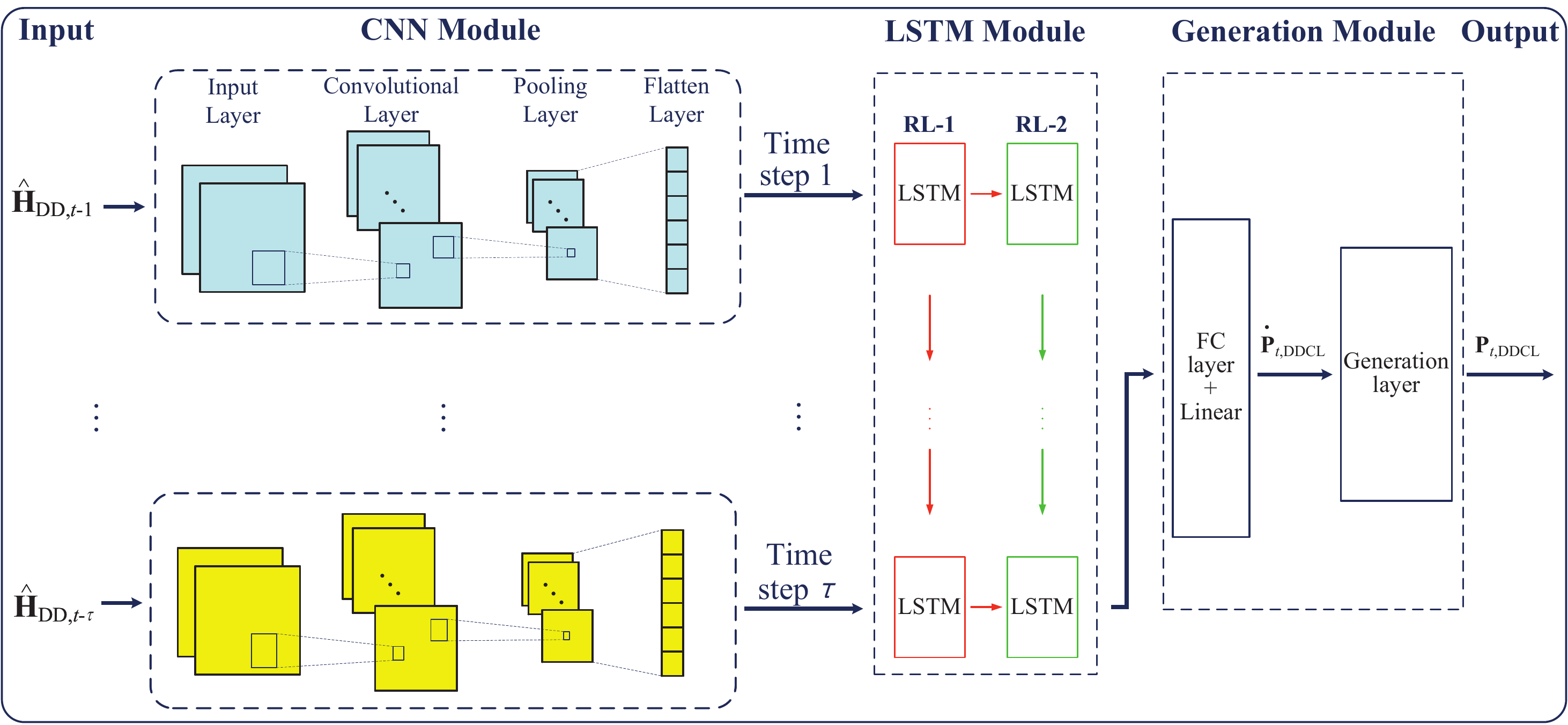}
  \caption{The proposed DDCL-Net architecture for predictive precoder design in the considered OTFS-enabled URLLC system.}\label{Fig:DDCL}
\end{figure*}

\section{Proposed DDCL-Net-based Predictive Precoder Design}

In this section, we adopt a powerful CLSTM architecture \cite{liu2019deep, lxm2020deepresidual} to develop a DDCL-Net to facilitate the spatial-temporal feature extraction from the historical channels for predictive precoder design.
In the following, we will introduce the details of the proposed DDCL-Net in terms of the neural network architecture and the associated precoder design algorithm.

\subsection{Architecture of the Proposed DDCL-Net}
As illustrated in Fig. \ref{Fig:DDCL}, the designed DDCL-Net is composed of the input, the CNN module, the LSTM module, the generation module, and the output.
Correspondingly, the hyperparameters of DDCL-Net is provided in Table \ref{Tab:Hyperparameters DDCL-Net}.
Each component of the DDCL-Net will be detailed in the following.

\begin{table}[t]
\normalsize
\caption{Hyperparameters of the Proposed DDCL-Net}\label{Tab:Hyperparameters DDCL-Net}
\centering
\small
\renewcommand{\arraystretch}{1.5}
\begin{tabular}{c c c}
  \hline
   \multicolumn{3}{l}{\textbf{Input}: $\tilde{\mathcal{H}}_{\mathrm{DD},{t}}^{\tau} \in \mathbb{R}^{\tau \times MN \times MN \times 2}$}  \\
  \hline
  \multicolumn{3}{l}{\textbf{CNN Module}:} \\
  \hspace{0.1cm} \textbf{Layers} & \textbf{Parameters} &  \hspace{0.3cm} \textbf{Values}   \\
  \hspace{0.1cm} Convolutional layer & Size of filters & \hspace{0.3cm}  $ 2 \times 3 \times 3 \times 2$   \\
  \hspace{0.1cm} Pooling layer & Size of filters & \hspace{0.3cm}  $ 2 \times 2 $   \\
  \hspace{0.1cm} Flatten layer & Output shape & \hspace{0.3cm}  $ 1024 \times 1 $   \\
  \multicolumn{3}{l}{\textbf{LSTM Module}:} \\
  \hspace{0.1cm} \textbf{Layers} & \textbf{Parameters} &  \hspace{0.3cm} \textbf{Values}   \\
  \hspace{0.1cm} RL-1 & Output shape & \hspace{0.3cm}  $ 32 \times \tau $   \\
  \hspace{0.1cm} RL-2 & Output shape & \hspace{0.3cm}  $ 32 \times 1 $   \\
  %\multicolumn{3}{l}{\textbf{Generation Module}:} \\
  %\hspace{0.1cm} \textbf{Layers} & \textbf{Parameters} &  \hspace{0.3cm} \textbf{Values}   \\
  %\hspace{0.1cm}  FC layer & Activation function & \hspace{0.3cm} Linear \\
  %\hspace{0.1cm}  FC layer & Output shape & \hspace{0.3cm} $ 2KMN $\\
  \hline
   \multicolumn{3}{l}{\textbf{Output}: $\mathbf{P}_{t,\mathrm{DDCL}} \in \mathbb{C}^{MN\times K}$}  \\
  \hline
\end{tabular}
\end{table}

According to the formulated problem $(\mathcal{P})$, the neural network input can be set as the historical channels from frame $t-1$ to frame $t-\tau$, which are successively processed at different time steps of the DDCL-Net.
For ease of processing, two independent neural network channels are leveraged to individually handle the real and imaginary parts of the input, i.e.,
\begin{equation}\label{input}
  \tilde{\mathcal{H}}_{\mathrm{DD},{t}}^{\tau} = \mathcal{M}([ \mathrm{Re}\{\mathcal{H}_{\mathrm{DD},{t}}^{\tau}\}, \mathrm{Im}\{\mathcal{H}_{\mathrm{DD},{t}}^{\tau}\} ]),
\end{equation}
where $\tilde{\mathcal{H}}_{\mathrm{DD},{t}}^{\tau}$ denotes the neural network input and  $\mathcal{M}(\cdot)\!:\mathbb{R}^{MN \times \tau MN}\mapsto\mathbb{R}^{\tau \times MN \times MN \times 2}$ is the mapping function.
Given the historical channels-based input, we then exploit feature extraction via the CNN module which consists of one input layer, one convolutional layer, one pooling payer, and one flatten layer.
In particular, each convolution is added with a rectified linear unit (ReLU) activation function and a maximum pooling is operated in the pooling layer.
Given the extracted features from the CNN module, we then adopt an LSTM module, which consists of two recurrent layers (RLs): RL-1 and RL-2, to further exploit the temporal dependency of the historical channels to implicitly predict the precoder for the next frame \cite{liu2022scalable}.
Specifically, since the output of the LSTM unit at the last time step in RL-2 fully exploits the temporal dependencies of historical channels, we adopt it as the final output of the entire LSTM model.
Subsequently, to generate the desired output, a generation module is added after the LSTM module.
In the generation module, a fully-connected (FC) layer with a linear activation function is first adopted to aggregate these extracted features from previous modules in a non-linear combination way.
Then, a generation layer with a power constraint-based normalization is adopted to generate the desired output.
In particular, the normalization process can be formulated as
\begin{equation}\label{P_t_normalization}
  \ddot{\mathbf{P}}_{t,\mathrm{DDCL}} = \sqrt{P_0}\frac{\dot{\mathbf{P}}_{t,\mathrm{DDCL}}}{\|\dot{\mathbf{P}}_{t,\mathrm{DDCL}}\|_F}
  \in \mathbb{C}^{2MN\times 1},
\end{equation}
where $\dot{\mathbf{P}}_{t,\mathrm{DDCL}} \in \mathbb{R}^{2MN\times K}$ denotes the real-valued output of the FC layer and $\ddot{\mathbf{P}}_{t,\mathrm{DDCL}}$ denotes the normalized result.
Thus, the output of the generation module is given by
\begin{equation}\label{P_t_generation}
\begin{aligned}
  &\mathbf{P}_{t,\mathrm{DDCL}} \\
  &= [\ddot{\mathbf{P}}_{t,\mathrm{DDCL}}(1:MN,:) + j\ddot{\mathbf{P}}_{t,\mathrm{DDCL}}(MN+1:2MN,:)],
\end{aligned}
\end{equation}
where $\ddot{\mathbf{P}}_{t,\mathrm{DDCL}}(1:MN,:)$ and $\ddot{\mathbf{P}}_{t,\mathrm{DDCL}}(MN+1:2MN,:)$ denote the real and imaginary parts of the designed precoder matrix.
Denote by $g_\omega(\cdot)$ the DDCL-Net with neural network parameters $\omega$, the output of the proposed DDCL-Net can be expressed as
\begin{equation}\label{}
  \mathbf{P}_{t,\mathrm{DDCL}} = g_\omega (\mathcal{H}_{{\mathrm{DD}},{t}}^{\tau}).
\end{equation}

\subsection{DDCL-Net-based Predictive Precoder Design Algorithm}
Based on the developed DDCL-Net, we then propose the DDCL-Net-based precoder design algorithm, which consists of the offline unsupervised training and the online precoder design.

\subsubsection{Offline Unsupervised Training}
Denote by
\begin{equation}\label{}
  \tilde{\mathcal{X}} \! = \! \left\{\!(\tilde{\mathcal{H}}_{{\mathrm{DD}},{t}}^{\tau(1)},\mathbf{H}_{{\mathrm{DD}},{t}}^ {(1)} ), (\tilde{\mathcal{H}}_{{\mathrm{DD}},{t}}^{\tau(2)},\mathbf{H}_{{\mathrm{DD}},{t}}^ {(2)} ), \! \cdots \! , \! (\tilde{\mathcal{H}}_{{\mathrm{DD}},{t}}^{\tau(N_t)},\mathbf{H}_{{\mathrm{DD}},{t}}^ {(N_t)} )\! \right\},
\end{equation}
the training set, where $N_t$ is the number of training examples and  $(\tilde{\mathcal{H}}_{{\mathrm{DD}},{t}}^{\tau(i)},\mathbf{H}_{{\mathrm{DD}},{t}}^ {(i)} )$ denotes the $i$-th, $i \in \{1,2,\cdots,N_t\}$, training example.
According to the problem formulation in (\ref{PF_objective}), the cost function can be designed as
\begin{equation}\label{cost_function_DDCL}
  J_{\mathrm{DDCL}}(\omega) = \frac{1}{N_t}\sum_{i=1}^{N_t}f_{\mathrm{FER}}\left(\mathbf{H}_{\mathrm{DD},{t}}^{(i)},g_\omega(\tilde{\mathcal{H}}_{{\mathrm{DD}},{t}}^{\tau(i)})\right).
\end{equation}
By exploiting the backpropagation (BP) algorithm, we can update the neural network parameters $\omega$ progressively to finally obtain the well-trained DDCL-Net, i.e.,
\begin{equation}\label{well_trained_DDCL}
  \mathbf{P}_{t,\mathrm{DDCL}}^{(i)} = g_{\omega^*} (\tilde{\mathcal{H}}_{{\mathrm{DD}},{t}}^{\tau(i)}), \forall i,
\end{equation}
where $g_{\omega^*}(\cdot)$ represents the well-trained DDCL-Net with the well-trained parameters $\omega^*$.

\subsubsection{Online Precoder Design}
Based on the well-trained DDCL-Net, we can operate the online phase, i.e., sending test examples to neural network to generate the predictive precoder matrix.
Given an arbitrary test example, denoted by $\tilde{\mathcal{H}}_{{\mathrm{DD}},{t}}^{\tau(i')}$, $i' \neq i$, the designed precoder can be expressed as
\begin{equation}\label{test_precoder}
  \mathbf{P}_{t,\mathrm{DDCL}}^{(i')} = g_{\omega^*} (\tilde{\mathcal{H}}_{{\mathrm{DD}},{t}}^{\tau(i')}).
\end{equation}

\subsubsection{Algorithm Steps}
Based on the above discussion, we propose the DDCL-Net-based predictive precoder design algorithm in Algorithm 1, where the number of iterations $N_{\mathrm{tr}}$ can be set based on the early stopping criteria \cite{goodfellow2016deep}.

\begin{table}[t]
\small
\centering
\begin{tabular}{l}
\toprule[1.8pt] \vspace{-0.3 cm}\\
\hspace{-0.1cm} \textbf{Algorithm 1} {DDCL-Net-based Predictive Precoder Design} \vspace{0.2 cm} \\
\toprule[1.8pt] \vspace{-0.3 cm}\\
\textbf{Algorithm Initialization:} $i_t = 0$ and the offline training set $\tilde{\mathcal{X}}$ \\
\textbf{Offline Unsupervised Training:} \\
1:\hspace{0.2cm}\textbf{Input:} Training set $\tilde{\mathcal{X}}$\\
2:\hspace{0.6cm}\textbf{while} $i_t \leq N_{\mathrm{tr}} $ \textbf{do} Updating $\omega$ \\
3:\hspace{0.8cm} to minimize $J_{\mathrm{DDCL}}(\omega)$ in (\ref{cost_function_DDCL}) based on BP algorithm  \\
\hspace{1.05cm} $i_t = i_t + 1$  \\
4:\hspace{0.6cm}\textbf{end while} \\
5:\hspace{0.2cm}\textbf{Output}:  Well-trained $g_{\omega^*}( \cdot ) $ as defined in (\ref{well_trained_DDCL})\\
\textbf{Online Precoder Design:} \\
6:\hspace{0.2cm}\textbf{Input:} Test data  $\tilde{\mathcal{H}}_{{\mathrm{DD}},{t}}^{\tau(i')}$  \\
7:\hspace{0.6cm} Calculate predictive precoder based on $g_{\omega^*}(\cdot)$ \\
8:\hspace{0.2cm}\textbf{Output:} $\mathbf{P}_{t,\mathrm{DDCL}}^{(i')} = g_{\omega^*} (\tilde{\mathcal{H}}_{{\mathrm{DD}},{t}}^{\tau(i')})$ \vspace{0.2cm}\\
\bottomrule[1.8pt]
\end{tabular}
\end{table}

\section{Simulation Results}
In this section, we conduct simulations for the considered OTFS-enabled URLLC, where both the transmitter and receiver are equipped with the single-antenna system.
Without loss of generality, we assume that the system is with $M = 8$, $N = 4$ and the carrier frequency and the sub-carrier spacing are set as 4 GHz and 15 kHz, respectively.
Moreover, the adopted DD domain channel can be generated according to (\ref{H_DD}) where the delay and Doppler indices follow the uniform distributions on $[0,l_{\max}]$ and $[-k_{\max},k_{\max}]$, respectively.
Correspondingly, we set $P=4$, $l_{\max}=5$, and the maximum speed of the mobile receiver is set as $100~\mathrm{km/h}$, which results in a $k_{\max} = 2$.
Also, in the considered channel offset model, we assume that $\epsilon_{\max} = -\epsilon_{\min} = 2$, $\varepsilon_{\max} = -\varepsilon_{\min} = 2$, and $\rho = 0.6$.
Furthermore, we focus on the sequential model with a $\tau = 5$ and an normalized mean squared error (NMSE) of $0.01$ is considered in the channel estimation.
To verify the effectiveness of the proposed method, we adopt three benchmark schemes for comparisons: (a) Lower bound scheme: A DL approach is adopted for precoder design with the perfect knowledge of ICSI at both the transmitter and receiver;
(b) MMSE equalizer-based scheme where we only adopt an MMSE equalizer \cite{tse2005fundamentals} at the receiver without the assistance of precoder;
(c) ZF equalizer-based scheme in which we only leverage an ZF equalizer \cite{tse2005fundamentals} at the receiver without the assistance of precoder.
The neural network parameters of the proposed DDCL-Net method are provided in Table \ref{Tab:Hyperparameters DDCL-Net}.
Moreover, for all the presented points in the results, if the achieved FER is at the level $10^{-\xi}$, more than $10^{\xi+1}$ Monte Carlo realizations are conducted for obtaining the average results \cite{rubinstein2016simulation, liu2014maximum}.
More detailed simulation results can be found in the journal version \cite{liu2023predictive} of this conference paper.

\begin{figure}[t]
  \centering
  \includegraphics[width=3.2in,height=2.8in]{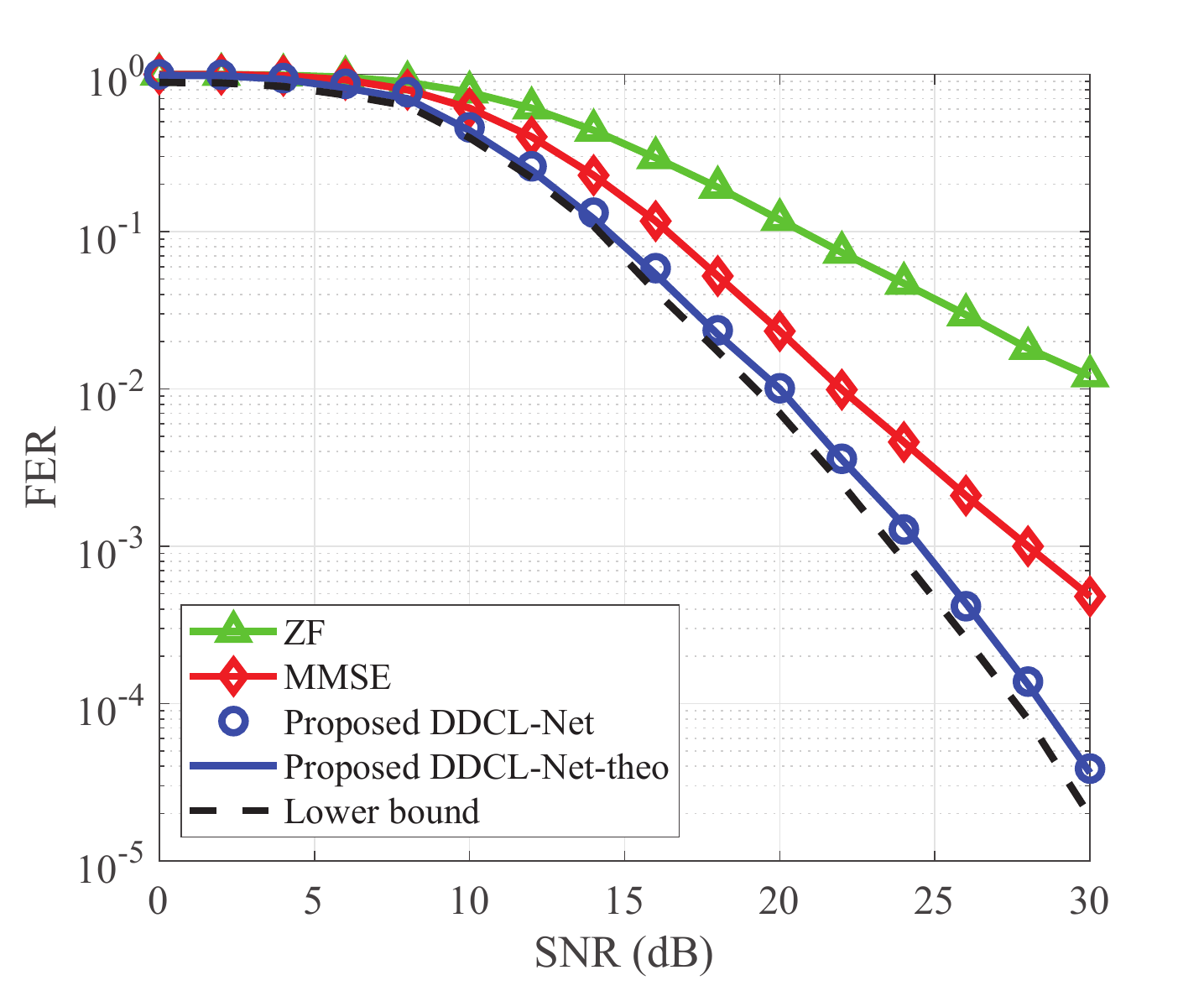}
  \caption{The curves of FER versus receive SNRs under QPSK modulation with $M=8$, $N=4$, and $K=32$.}\label{Fig:FER_SNR}
\end{figure}

Fig. \ref{Fig:FER_SNR} investigates the FER performance versus the received signal-to-noise ratio (SNR) under $K=MN$ with quadrature phase shift keying (QPSK) modulation.
We can first observe that for the proposed method, the theoretical results (denoted by ``DDCL-Net-theo'') match very well with its corresponding simulation results (denoted by ``DDCL-Net''), which proves the efficiency of the derived FER expression in (\ref{FER}).
Also, it is observed that for all the considered algorithms, the FER values decrease with the increase of SNR.
This is as expected since a large SNR value indicates a stronger received signal power.
In addition, the MMSE method achieves a relatively better performance than the ZF method. However, without the assistance of the precoder design, both these two methods have large performance gap compared with the lower bound scheme.
In contrast, our proposed method significantly outperforms the ZF and MMSE methods, achieving almost the same FER performance as the lower bound scheme.
This is because our proposed method can leverage the historical channels to learn the temporal dependency to implicitly predict the precoder, which further improves the system performance.

\begin{figure}[t]
  \centering
  \includegraphics[width=3.2in,height=2.8in]{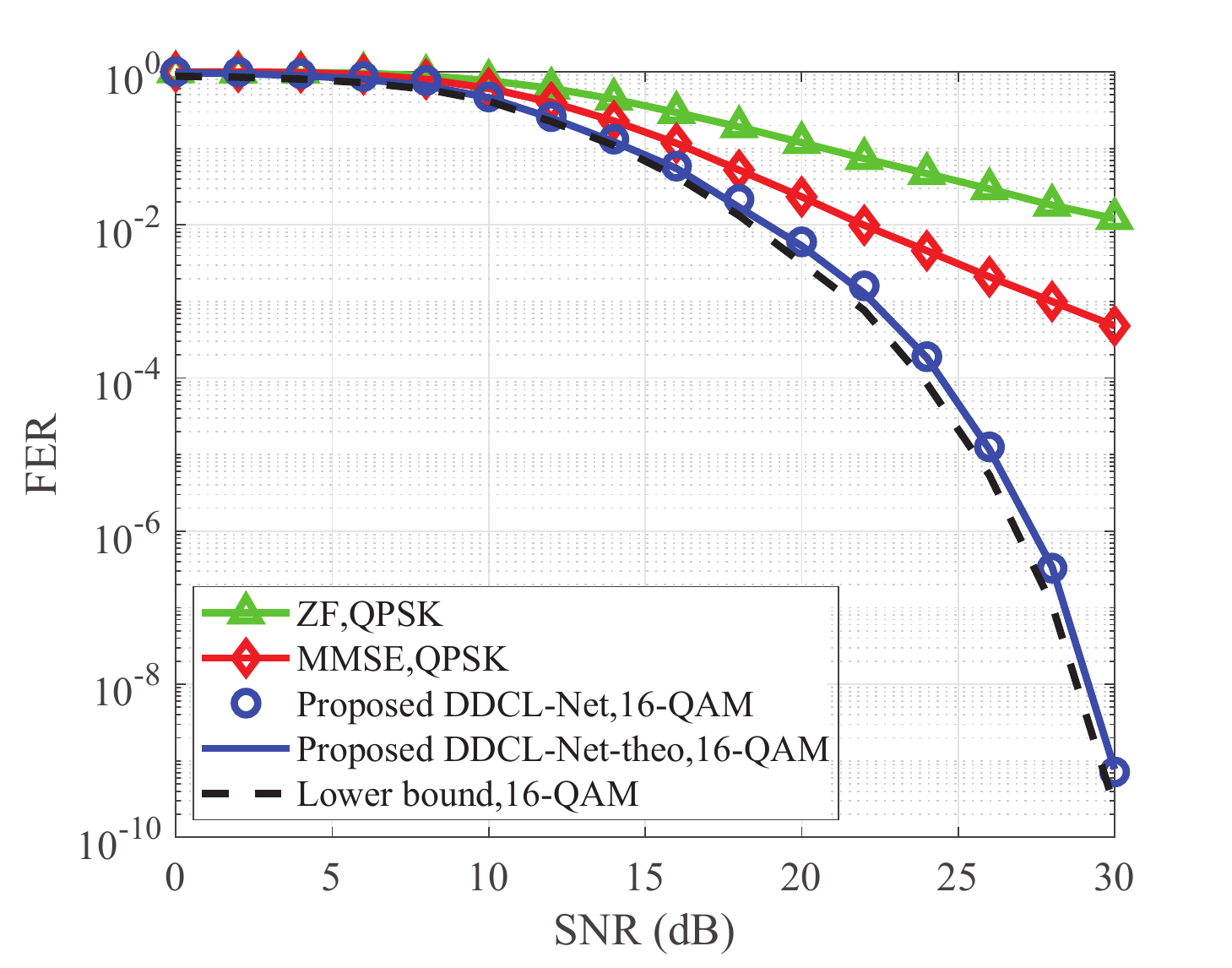} 
  \caption{The curves of FER versus receive SNRs with $M=8$, $N=4$, and $K=16$.}\label{Fig:FER_SNR_QAM}\vspace{-0.1cm}
\end{figure}

Correspondingly, Fig. \ref{Fig:FER_SNR_QAM} also studies the FER performance versus the received SNR in the case of $K < MN$, i.e., $K=\frac{1}{2}MN$.
In particular, a 16-quadrature amplitude modulation (QAM) is adopted in our proposed method to guarantee the same data rate with that of the QPSK scheme in the ZF and MMSE methods.
Similar to the results in Fig. \ref{Fig:FER_SNR}, the performance of our proposed method is obviously superior to that of the ZF and MMSE methods.
In particular, our proposed method can even achieve an FER value of less than $10^{-9}$ when $\mathrm{SNR}=30~\mathrm{dB}$, which guarantees an ultra reliability for data transmission.
The reason is that the precoder design with $K < MN$ indicates a longer channel coding on the source symbols, which facilitates the FER performance improvement.

\section{Conclusion}
In this paper, we adopted a DL approach for predictive precoder design in OTFS-enabled URLLC.
We first formulated a universal predictive precoder design problem where a closed-form theoretical FER expression was derived as the objective function to characterize the system reliability.
Then, to address the formulated problem, a DDCL-Net was developed in which both the CNN and LSTM
modules were successively adopted to exploit the spatial-temporal features of wireless channels to facilitate the predictive task. Finally, simulation results demonstrated that our proposed method can achieve almost the same performance as the lower bound method which is conducted with the availability of perfect ICSI at both the transmitter and receiver. %\cite{liu2022learning} \cite{liu2020deeptransfer} \cite{liu2022deepresidual} \cite{liu2019deep}
%\cite{lxm2020deepresidual}  \cite{liu2022scalable} \cite{li2022novel}
%\cite{wang2017deep} \cite{o2017introduction} 
%\cite{hu2021robust}  \cite{cai2022resource} \cite{li2022many} 
%\vspace{0.2cm}

\bibliographystyle{ieeetr}

\setlength{\baselineskip}{10pt}

\bibliography{ReferenceSCI2}

\end{document}